\documentstyle[11pt,newpasp,twoside,epsf]{article}
\def\edcomment#1{\iffalse\marginpar{\raggedright\sl#1\/}\else\relax\fi}
\reversemarginpar

\begin{document}
\title{Holes and high velocity H{\small\ I} in NGC 6946}
\author{Rense Boomsma, Thijs van der Hulst}
\affil{Kapteyn Astronomical Institute, P.O. Box 800, 9700 AV Groningen, the Netherlands}
\author{Tom Oosterloo, Filippo Fraternali}
\affil{ASTRON, P.O. Box 2, 7990 AA Dwingeloo, the Netherlands}
\author{Renzo Sancisi}
\affil{Osservatorio Astronomico, Via Ranzani 1, 40127 Bologna, Italy}
\affil{Kapteyn Astronomical Institute, Groningen, the Netherlands}

\begin{abstract}
We are studying the properties of the holes and the high velocity gas in NGC 6946. Here we present some puzzling results.
\end{abstract}

\section{Introduction}
In several spiral galaxies H{\footnotesize\ I} has been detected in the halo. In NGC 6946 Kamphuis \& Sancisi (1993) discovered widespread gas with velocities deviating strongly from galactic rotation. Studies of the edge-on spiral galaxy NGC 891 (Swaters et al., 1997) and the inclined spiral galaxy NGC 2403 (Fraternali et al. 2001, Schaap et al. 2000) revealed a thick layer of H{\footnotesize\ I} rotating more slowly than the thin disk. This has led to the picture of spiral galaxies with a halo population of high velocity H{\footnotesize\ I}. A few explanations have been proposed for this phenomenon, one of which is the galactic fountain mechanism with massive star formation and supernovae blowing the H{\footnotesize\ I} into the halo.\\
To investigate further the 3D picture of the anomalous H{\footnotesize\ I}, we have observed \mbox{NGC 6946} with very high sensitivity with the Westerbork Synthesis Radio Telescope. \mbox{NGC 6946} is a large nearby spiral galaxy seen almost face-on and is known for its high star-formation rate.\\

\section{Holes and high velocity gas}
The H{\footnotesize\ I} disk of \mbox{NGC 6946} shows a large number of holes (see fig. 1, top left panel) and gas with velocities deviating 50 to 100 km/s from normal galactic rotation. The high velocity H{\footnotesize\ I} is widespread and mostly located in the inner (bright optical) part of the disk (see optical image in fig. 1), suggesting a close relation with the ongoing star formation and supernovae. The gas is probably brought into the halo by galactic fountains.\\
\begin{figure}[h]
\plotone{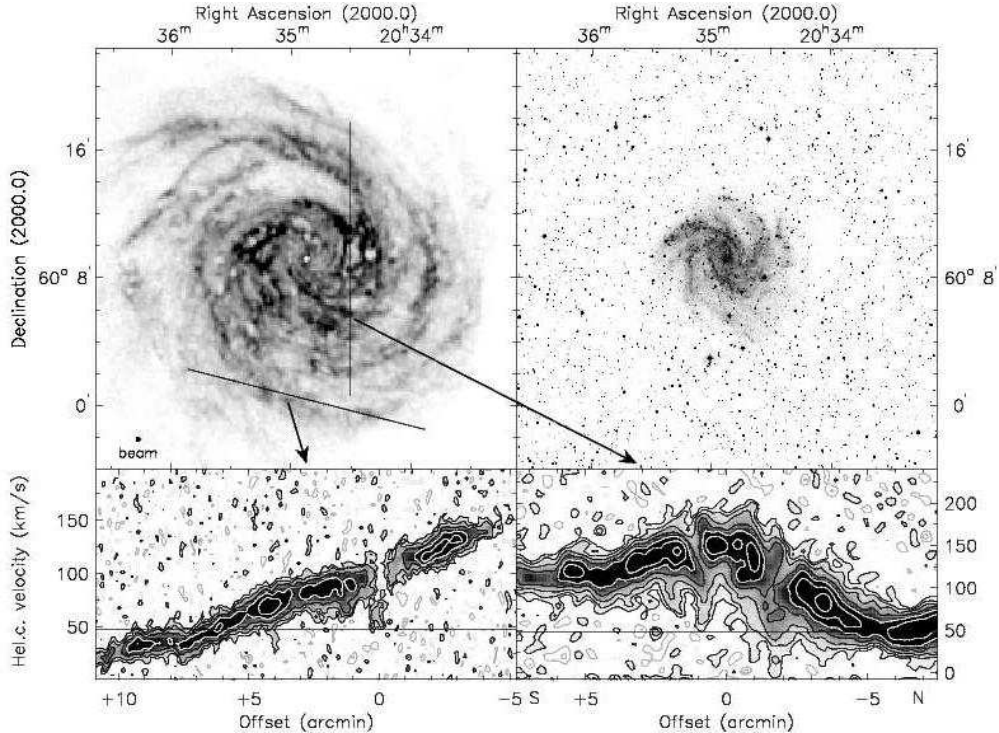}
\caption{The top left panel shows the total H{\footnotesize\ I} map of NGC 6946; the top right panel shows a DSS image of the same region on the same scale, the bottom two panels show position-velocity diagrams taken along the lines drawn in the H{\footnotesize\ I} map.}
\end{figure}
In many cases in the inner disk there is a clear association between the high velocity gas and the holes (fig. 1, bottom right panel) where also many H{\footnotesize\ II} regions are located. However, high velocity H{\footnotesize\ I} and holes are also found in the outer H{\footnotesize\ I} layer of \mbox{NGC 6946} beyond the optical disk (see fig. 1, bottom left panel), where very little or no star formation is present. Star formation is unlikely to be the cause, so another mechanism is required. Perhaps this is the result of infalling clouds onto the galaxy disk.
\acknowledgments{The Westerbork Synthesis Radio Telescope is operated by the Netherlands Foundation for Radio Astronomy with financial support from the Netherlands Foundation for the Advancement of Pure Research (N.W.O)}

\end{document}